\documentclass[twocolumn,notitlepage,nofootinbib,amsmath,amssymb,aps,prl]{revtex4-1}

\usepackage[T1]{fontenc}
\usepackage{graphicx}% Include figure files
\usepackage{dcolumn}% Align table columns on decimal point
\usepackage{bm}% bold math
\begin{document}

\title{Learning Bayesian posteriors with neural networks for gravitational-wave inference}

\author{Alvin J.\ K.\ Chua}
\email{Alvin.J.Chua@jpl.nasa.gov}
\affiliation{Jet Propulsion Laboratory, California Institute of Technology, Pasadena, CA 91109, U.S.A.}

\author{Michele Vallisneri}
\email{Michele.Vallisneri@jpl.nasa.gov}
\affiliation{Jet Propulsion Laboratory, California Institute of Technology, Pasadena, CA 91109, U.S.A.}

%\author{Chad R.\ Galley}
%\affiliation{Jet Propulsion Laboratory, California Institute of Technology, Pasadena 91109 USA}

\date{\today}

\begin{abstract}
We seek to achieve the Holy Grail of Bayesian inference for gravitational-wave astronomy: using deep-learning techniques to instantly produce the posterior $p(\theta|D)$ for the source parameters $\theta$, given the detector data $D$. To do so, we train a deep neural network to take as input a signal + noise data set (drawn from the astrophysical source-parameter prior and the sampling distribution of detector noise), and to output a parametrized approximation of the corresponding posterior. We rely on a compact representation of the data based on reduced-order modeling, which we generate efficiently using a separate neural-network waveform interpolant [A.\,J.\,K.\,Chua,\,C.\,R.\,Galley\,\&\,M.\,Vallisneri, Phys.\,Rev.\,Lett.\,\textbf{122},\,211101\,(2019)]. Our scheme has broad relevance to gravitational-wave applications such as low-latency parameter estimation and characterizing the science returns of future experiments. Source code and trained networks are available online at \url{github.com/vallis/truebayes}.
\end{abstract}

%\pacs{Valid PACS appear here}
\maketitle

%\tableofcontents

\paragraph{Introduction.}

In the Bayesian analysis of signals immersed in noise \cite{gregory2005bayesian}, we seek a representation for the posterior probability of one or more parameters that govern the shape of the signals. Unless the parameter-to-signal map (the \emph{forward model}) is very simple, the analysis (or \emph{inverse solution}) comes at significant computational cost, as it requires the stochastic exploration of the likelihood surface over parameter space. Such is the case, for instance, of parameter estimation for gravitational-wave sources such as the compact binaries detected by LIGO--Virgo \cite{ligo2018gwtc,abbott2019binary}; here each likelihood evaluation requires that we generate the gravitational waveform corresponding to a set of source parameters, and compute its noise-weighted correlation with detector data \cite{creighton2011gravitational}. Waveform generation is usually the costlier operation, so gravitational-wave analysts often utilize faster, less accurate waveform models \cite{PhysRevLett.113.151101,PhysRevD.89.084006}, or accelerated \emph{surrogates} of slower, more accurate models \cite{PhysRevLett.115.121102}.

Extending the analysis from the data we have obtained to the data we might expect (i.e., scoping out parameter estimation for future experiments) further compounds the expense, since we need to explore posteriors for many noise realizations, and across the domain of possible source parameters. For concreteness, let us price the evaluation of a single Bayesian posterior at $\gtrsim 10^6$ times the cost of generating a waveform, and the characterization of parameter estimation for a single source type at $\gtrsim 10^6$ times the cost of a posterior. With current computational resources, this means that (for instance) accurate component-mass estimates only become available hours or days after the detection of a binary black-hole coalescence \cite{Usman_2016,PhysRevD.95.042001}, while any extensive study of parameter-estimation prospects must rely on less reliable techniques such as the Fisher-matrix approximation \cite{PhysRevD.77.042001}.

In this Letter, we show how one- or two-dimensional projections of Bayesian posteriors may be produced using \emph{deep neural networks} \cite{Goodfellow-et-al-2016} trained on large ensembles of signal + noise data streams. (Specifically, we adopt \emph{multilayer perceptrons} \cite{haykin1999neural}, although other architectures are likely to be viable.) The network for each source parameter or parameter pair takes as input a noisy signal, and instantly outputs an approximate posterior, represented either as a histogram or parametrically (e.g., as a Gaussian mixture). We dub such networks \textsc{Percival}:\footnote{After the original achiever of the Grail \cite{10.2307/j.ctt1npbmx}.} Posterior Estimation Results Computed Instantaneously Via Artificial Learning. Crucially, the \emph{loss function} used in the training of these networks does not require that we compute the likelihood or posterior for each training signal, but only that we provide the corresponding source parameters. If the training set reflects the assumed prior distribution of the parameters and sampling distribution of the noise, the resulting posteriors approximate the correct Bayesian distribution---and indeed achieve it asymptotically, for very large training sets and networks.

In the gravitational-wave context, the training of a \textsc{Percival} network can require up to $\sim$ $10^9$ waveform evaluations, which is $\lesssim10^3$ times the cost of stochastically exploring the posterior for a single noisy signal. However, the costly training is performed \emph{offline}; afterwards, the networks can perform inference on multiple signals with negligible execution times. The prompt generation of posterior parameter distributions for binary coalescence alerts \cite{Abbott_2019} could be a worthy use of this speed (once suitable networks are trained, which we do not attempt here). Another potential application is the generation of effective proposal kernels to facilitate more detailed posterior analyses with Markov chain Monte Carlo methods \cite{PhysRevD.58.082001}. Last, very rapid inference could also prove useful in the characterization of parameter-estimation prospects for next-generation detectors such as LISA \cite{2017arXiv170200786A}.

The still somewhat hefty cost of training \textsc{Percival} can be offset significantly by pairing it with its forward counterpart \textsc{Roman} \cite{PhysRevLett.122.211101}, which is essentially a neural network that has been fitted to the relevant waveform model in a \emph{reduced-basis} representation \cite{PhysRevLett.106.221102}. \textsc{Roman} (Reduced-Order Modeling with Artificial Neurons) takes as input a set of source parameters, and outputs in milliseconds the corresponding signal at high accuracy; it provides a conceptually cleaner alternative to the combined analysis framework comprising surrogate waveforms \cite{PhysRevX.4.031006} and the likelihood compression technique known as \emph{reduced-order quadrature} \cite{PhysRevD.87.124005}. Being a neural network, \textsc{Roman} has additional features such as analytic waveform derivatives and---more pertinently for this work---the generation of waveforms in large batches at next to no marginal cost. The latter allows us to fully exploit highly parallel GPU architectures to build training sets for \textsc{Percival} that are effectively infinite in size, which in turn grants immunity to the perennial deep-learning problem of overfitting.

Deep-learning techniques have gained popularity in the gravitational-wave community over the past two years, with the majority of efforts focused on applying \emph{convolutional neural networks} \cite{LeCun:1998:CNI:303568.303704} to the \emph{classification} task of signal detection, specifically for transient signals from compact binaries in ground-based detector data \cite{GebKilParHarSch17,PhysRevLett.120.141103,PhysRevD.97.044039,GEORGE201864,Fan2019,PhysRevD.99.124032,PhysRevD.100.044025,shen2019deep,gebhard2019convolutional,krastev2019real}. They are also being investigated as detection tools for persistent signals from asymmetric neutron stars \cite{PhysRevD.99.024024,PhysRevD.100.044009,morawski2019deep,2019arXiv190902262M}. While the application of neural networks to the \emph{regression} task of source parameter estimation is addressed in some of these papers \cite{PhysRevD.97.044039,Fan2019,PhysRevD.99.124032,GEORGE201864}, there it is restricted to the recovery of pointwise estimates, and with a frequentist characterization of errors based only on the test set. However, near the completion of this manuscript, we learned of related work by Gabbard et al. \cite{gabbard}, where a conditional variational autoencoder \cite{tonolini2019variational} is trained on parameter--signal pairs to output samples from the Bayesian posterior. We expect that comparison between our method and theirs will offer useful insight.

\paragraph{Training neural networks to produce posteriors.} We now describe our scheme to perform Bayesian posterior estimation using neural networks. A \emph{perceptron classifier} is a network that takes a data vector $D$ as input and outputs the estimated probabilities $q_i[D]$ that the input belongs to each member of a universal set of $N$ disjoint classes $\{C_i\}$. The \emph{Bayesian optimal discriminant} is the classifier that returns the posterior probabilities $p_i[D] := p(C_i|D) \propto p(D|C_i) \, p(C_i)$, where $p(D|C_i)$ is the likelihood of $D$ occurring in $C_i$, and $p(C_i)$ is the prior probability of $C_i$ itself. It is a well-established result in the machine-learning literature that perceptron classifiers can approximate Bayesian optimal discriminants \cite{ruck1990multilayer,wan1990neural} when they are trained on a population of inputs $\{ D_j \}$ distributed according to $p(D,C_i) = p(D|C_i) \, p(C_i)$. This is achieved by minimizing (over the network weights) the loss function $\sum_j \lambda(\mathbf{1}_i[D_j],q_i[D_j])$, where $\mathbf{1}_i$ is the indicator function of $C_i$, and $\lambda$ is some vector distance on the space $[0,1]^N$ (e.g., the squared $\ell^2$ norm, or the discrete Kullback--Leibler divergence \cite{pollard2002user}).

Note that the above training procedure does not require explicit evaluation of the values $p(D_j|C_i)$ or $p_i[D_j]$, but only a \emph{generative model} of the data, i.e., the ability to randomly draw classes $C_i$ from the prior $p(C_i)$ and data vectors $D_j$ from the conditional sampling distribution $p(D|C_i)$. To move from classifiers to posterior densities for continuous parameters, we may simply define classes based on a binning of the parameter domain of interest, in which case the network will output histograms of the target posterior. This coarse-graining of parameter space highlights the relationship between classification and regression, but is not actually necessary for our scheme, since the network can instead output a parametric posterior representation (e.g., a Gaussian mixture) to be fed into an analogous loss function. Although histograms might be impractical for posterior projections in $>2$ dimensions, the parametric approach extends more readily to higher-dimensional projections (which are seldom used for visualization and interpretation in any case).

Let us now explicitly derive the loss function used in our scheme. Consider a data model described by the continuous parameters $\vartheta:=\{\theta,\phi\}$, with $\theta$ denoting the parameters for which we seek a posterior. We want to minimize the statistical distance $\lambda(p,q)$ between the (marginalized) true posterior $p(\theta|D)=\int \mathrm{d}\phi\,p(\vartheta|D)$ and the network-estimated posterior $q(\theta|D)$, integrated over the distribution of the data $p(D)$.
% This might only be attained if the training set and network capacity are idealized (possibly infinite), but can nevertheless be approached by minimizing the quantity $\Lambda:=\int\mathrm{d}D\,p(D)\,\lambda$ with respect to the network weights.
If $\lambda$ is the Kullback--Leibler divergence, we have the integrated distance
\begin{equation}
\label{eq:lambda}
\int \mathrm{d}D \, p(D)\left[\int \mathrm{d}\theta \, p(\theta|D) \ln{\tfrac{p(\theta|D)}{q(\theta|D)}} \right].
\end{equation}
Using Bayes' theorem on $p(\vartheta|D)$ and dropping the term containing $p(D,\vartheta)\ln{p(\theta|D)}$ (which is constant with respect to the network weights), we obtain the loss function
\begin{equation}
\label{eq:loss}
L := -\iint \mathrm{d}D \, \mathrm{d}\vartheta \, p(D,\vartheta) \ln{q(\theta|D)},
\end{equation}
which can be approximated (modulo normalization) as a discrete sum
%\footnote{To understand how the double integral in Eq.\ \eqref{eq:loss} is approximated by Eq.\ \eqref{eq:sumoverbatch}, consider the model $D = f(x) + n$, with $f$ a deterministic function and the noise $n \sim p(n)$; in that case $p(D|x) = p(n - f(x))$, so integrating over $D$ is equivalent to integrating over noise realizations.}
over a notional training batch $\{(\vartheta_j,D_j)\}$:
\begin{equation}
\label{eq:sumoverbatch}
L \simeq -\sum_j \ln{q(\theta_j|D_j)},
\end{equation}
where, crucially, each $\vartheta_j$ is first drawn from the prior $p(\vartheta)$ before $D_j$ is drawn from the conditional $p(D|\vartheta_j)$.

The summand in Eq.\ \eqref{eq:sumoverbatch} is precisely the $q$-dependent term in the Kullback--Leibler divergence between the Dirac delta function $\delta(\theta - \theta_j)$ and $q(\theta|D_j)$. Likewise, if the network-estimated posterior is represented as a histogram $q_i[D]$, Eq.\ \eqref{eq:sumoverbatch} simplifies to 
\begin{equation}
\label{eq:crossentropy}
L \simeq -\sum_{ij} \mathbf{1}_i[\theta_j]\ln{q_i[D_j]},
\end{equation}
where $\mathbf{1}_i$ is now the indicator function of the $i$-th histogram bin. Eq.\ \eqref{eq:crossentropy} is familiar to machine-learning practitioners, and is more commonly known as the cross-entropy loss for classification problems \cite{Goodfellow-et-al-2016}. Finally, note that a derivation similar to the one above can also be given for the squared $\ell^2$ distance $\int \mathrm{d}\theta \, |p(\theta|D) - q(\theta|D)|^2$, resulting in an alternative (but less tractable) loss
\begin{equation}
\label{eq:l2loss}
L' \simeq -\sum_j \left[
2 q(\theta_j|D_j) - \int \mathrm{d}\theta \, |q(\theta|D_j)|^2 \right].
\end{equation}

\paragraph{Leveraging reduced waveform representations.}

In gravitational-wave astronomy, the data $D$ is usually a time or frequency series of strain $h$ measured by the detector. For the transient signals observed by ground-based detectors, $h$ is typically a vector of length $\lesssim 10^4$; this rises to $\lesssim 10^8$ for persistent LIGO--Virgo signals, or the mHz-band signals sought by future space-based detectors. Once the presence of a signal is established, Bayesian inference proceeds via the canonical likelihood
\begin{equation}
\label{eq:gwlikelihood}
p(D|\theta)\propto\exp{\left\{-\tfrac{1}{2}\langle h(\vartheta)-D|h(\vartheta)-D\rangle\right\}},
\end{equation}
where $\langle\cdot|\cdot\rangle$ is a noise-weighted inner product that incorporates a power spectral density model for the detector noise (assumed to be Gaussian and additive).

As mentioned earlier, the generation of the waveform template $h(\vartheta)$ and the evaluation of the inner product are both computationally expensive, due to the complexity of relativistic modeling and the large dimensionality of the inner-product space. In the \emph{reduced-order modeling} framework \cite{PhysRevLett.106.221102}, we mitigate this cost by constructing a linear reduced basis for the (far more compact) template manifold embedded in the inner-product space, as well as a fast interpolant for the template model in this reduced representation. The neural network \textsc{Roman} \cite{PhysRevLett.122.211101} implements such an interpolation, allowing Eq.\ \eqref{eq:gwlikelihood} to be cast in the reduced but statistically equivalent form \cite{PhysRevLett.122.211101}
\begin{equation}\label{eq:romanlikelihood}
p(\beta|\theta)\propto\exp{\left\{-\tfrac{1}{2}\left|\alpha(\vartheta)-\beta\right|^2\right\}},
\end{equation}
where the $d$-vectors $\alpha(\vartheta)$ and $\beta$ are the projections of $h(\vartheta)$ and $D$, respectively, onto the reduced basis; furthermore, $\beta=\alpha(\vartheta_*)+\nu$ for the true source parameters $\vartheta_*$ and the (projected) noise realization $\nu\sim\mathcal{N}(0,I_d)$.

In this Letter, we give a demonstration of \textsc{Percival} by using the four-parameter, $d \simeq 200$ \textsc{Roman} network described fully in \cite{PhysRevLett.122.211101}, which embodies the family of 2.5PN TaylorF2 waveforms \cite{PhysRevD.79.104023} emitted by inspiraling black-hole binaries with aligned spins $\chi_{1,2}\in[-1,1]$ and component masses $m_{1,2}\in[1.25,10]\times10^5M_\odot$.
%It is constructed from a 2.5PN TaylorF2 waveform model \cite{PhysRevD.79.104023} over the frequency range $[1,10]\,\mathrm{mHz}$, and represents the $d\approx200$ space of possible signals from an inspiraling black-hole binary with aligned spins $\chi_{1,2}\in[-1,1]$ and component masses $m_{1,2}\in[1.25,10]\times10^5M_\odot$.
% This corresponds to a subset of the high-redshift massive-black-hole binaries that will be observed by LISA \cite{2017arXiv170200786A}.
The \textsc{Roman} templates are normalized to unit signal-to-noise ratio $\rho:=|\alpha|=\sqrt{\langle h|h\rangle}$. Here, we vary signal amplitudes by setting $\alpha\to\rho\alpha$; our model parameters are then $\vartheta=(M_c,\eta,\chi_1,\chi_2,\rho)$, where the chirp mass $M_c$ and symmetric mass ratio $\eta$ are an alternative parametrization of the $m_{1,2}$ space. Large batches of \textsc{Roman} signals, with noise added as described after Eq.\ \eqref{eq:romanlikelihood}, can be generated in $\sim 0.1 \,\mathrm{s}$ on a 2014 Tesla K80 GPU.

% For the training of \textsc{Percival}, we use batches of $10^5$ signals that are distributed according to simple uniform priors in parameter space, and ensure that the network is shown a different batch in each training epoch to eliminate overfitting.

\paragraph{Example results.}

\begin{figure}[!tbp]
\centering
\includegraphics[width=\columnwidth]{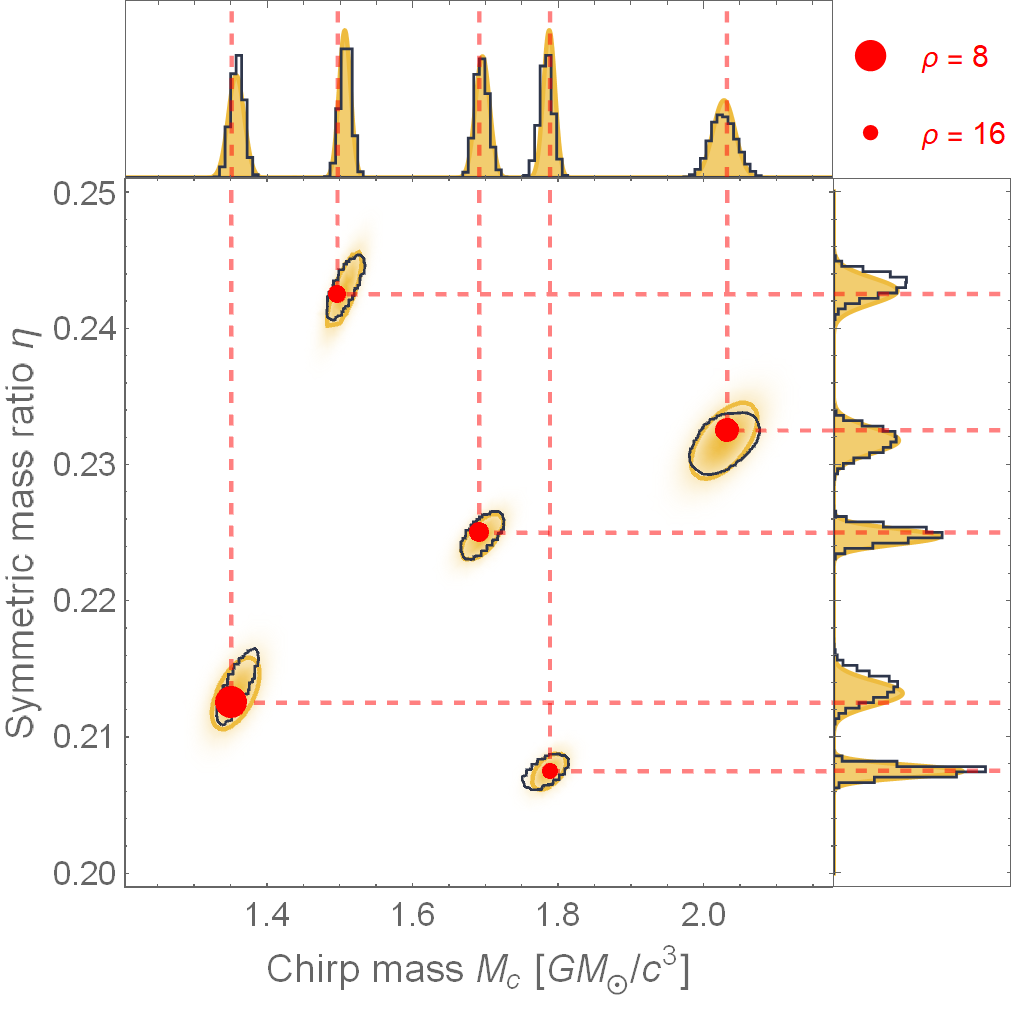}
\caption{Estimated (yellow) and true (black) posteriors for five test signals (red), with parameters of interest $(M_c,\eta)$ and the prior $p_1(\vartheta)$. Contours in main panel are three-sigma-equivalent. Posteriors in side panels are further marginalized over each parameter. Unit of $M_c$ is one Solar mass in seconds.}
\label{fig:2d}
\end{figure}

The posteriors on the full domain of the \textsc{Roman} model are nontrivial: they are highly multimodal at sub-threshold signal-to-noise ratios $\rho<8$, and this persists even for $\rho\in[8,16]$ due to correlations in the $(\eta,\chi_1,\chi_2)$ space (although $M_c$ is then better constrained). Ironically, this results from the simplicity of the TaylorF2 waveform used in the initial \textsc{Roman} demonstration; work is underway to build a more realistic inspiral--merger--ringdown \textsc{Roman} model, which will in turn admit less degenerate posteriors. Another issue is that the parameter space is quite large from a Fisher-information perspective, resulting in posteriors that are highly localized and hence difficult to resolve. We train a number of one- and two-dimensional \textsc{Percival} networks with different priors but, for the above reasons, obtain acceptable convergence to the true posteriors only when the priors are confined to smaller subspaces.

Our first successful example is a network that estimates the joint posterior of the parameters $\theta=(M_c,\eta)$, for signals distributed according to the uniform prior
\begin{equation}
\label{eq:prior1}
p_1(\vartheta)\propto\mathbf{1}_{\Delta M_c\times\Delta\eta}(M_c,\eta)\,\delta(\chi_1)\,\delta(\chi_2)\,\mathbf{1}_{\Delta\rho}(\rho),
\end{equation}
where the prior intervals are given by $\Delta M_c=[2.4,4.5]\times10^5M_\odot$, $\Delta\eta=[0.2,0.25]$ and $\Delta\rho=[8,16]$. This is effectively a non-spinning three-parameter submodel. Our \textsc{Percival} network takes as input signal + noise coefficients $\beta$, processes them through eight hidden layers of width 1024 (with the \emph{leaky ReLU} activation function \cite{Maas2013RectifierNI} applied to each), and outputs (with linear activation) a vector of five quantities that specify a single bivariate Gaussian. Training is performed using batch gradient descent with \emph{Adam optimization} \cite{2014arXiv1412.6980K} and a manually decayed learning rate; to eliminate overfitting, we generate a new batch of $10^5$ signals at every training epoch.

The network is fed $\sim 10^9$ examples before the loss levels off. While this figure might appear excessive for the $\leq5$-parameter examples in this work, note that it only arises from our choice to leverage the sheer availability of \textsc{Roman} templates, and does not apply to the traditional approach of training on a much smaller set (but then having to guard against overfitting). The number of required training examples is also unlikely to scale exponentially with the number of model parameters, since the complexity of the posterior space is determined more strongly by the information content of the waveform model and the degree of degeneracy among its parameters---neither of which are necessarily tied to dimensionality.
\begin{figure}[!tbp]
\centering
\includegraphics[width=\columnwidth]{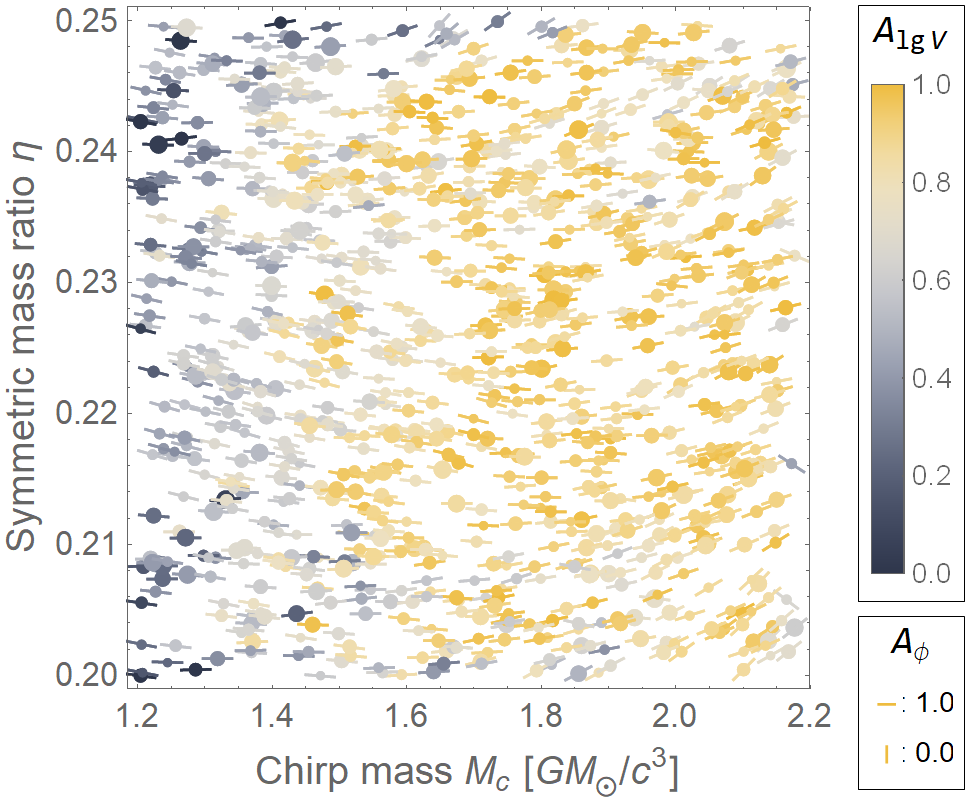}
\caption{Accuracy of estimated covariance matrix in terms of associated ellipse volume ($A_{\lg{V}}$) and orientation ($A_\phi$) for 1000 test signals, using the network of Fig.\ \ref{fig:2d}. A value of 1 for $A_{\lg{V}}$ corresponds to exact prediction, and 0 to errors of more than $\times 10$; likewise, a value of 1 for $A_\phi$ is exact prediction, and 0 is the maximal angular error of $\pi/2$. The radius of each point is inversely proportional to the signal-to-noise ratio.}
\label{fig:cov}
\end{figure}

\begin{figure}[!tbp]
\centering
\includegraphics[width=\columnwidth]{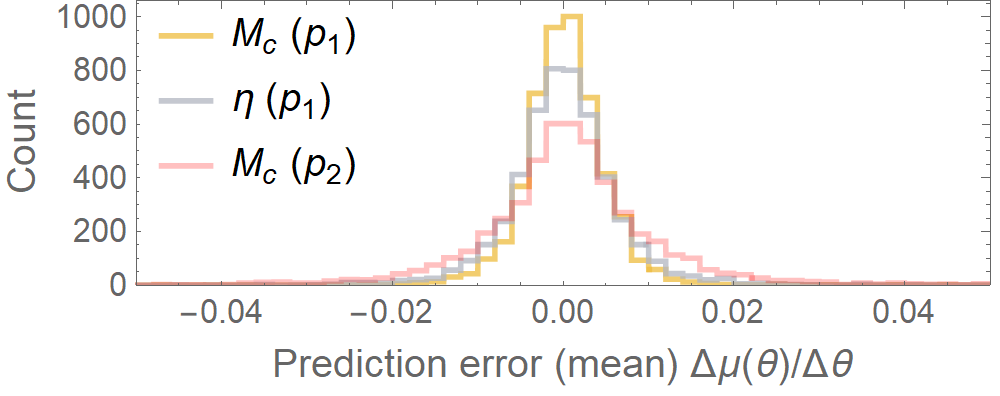}
\includegraphics[width=\columnwidth]{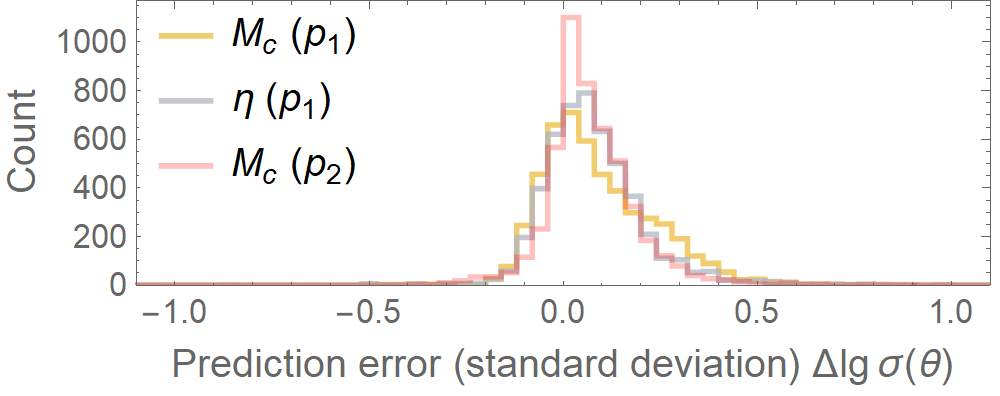}
\caption{Prediction error in mean values (top) and standard deviations (bottom) of posteriors for 5000 test signals, using the network of Fig.\ \ref{fig:2d}. Also included are results for an $M_c$-only network that is trained and tested on the prior $p_2(\vartheta)$.}
\label{fig:meanstd}
\end{figure}

In Fig.\ \ref{fig:2d}, we compare the \textsc{Percival}-estimated posteriors for five test signals against the corresponding true posteriors, which are computed using a brute-force oversampling approach that exploits the batch-generation capability of \textsc{Roman}. The trained network localizes the posteriors well, and captures their near-Gaussian correlation structure with reasonable accuracy. This statement is made more quantitative in Figs\ \ref{fig:cov} and \ref{fig:meanstd}, where the performance of the network is assessed on larger test sets. Fig.\ \ref{fig:cov} depicts the accuracy of the network-estimated posterior covariance matrix $\Sigma$, for 1000 test signals distributed according to $p_1(\vartheta)$. Each point on the plot corresponds to a single matrix, with color describing the accuracy of the associated ellipse $2$-volume $V:=\sqrt{\det{\Sigma}}$:
\begin{equation}
A_{\lg{V}}:=\max{\left\{1-\left|\lg{\tfrac{V}{V_\mathrm{true}}}\right|,0\right\}},
\end{equation}
where $V_\mathrm{true}$ is obtained from the sample covariance matrix of the true posterior. The angle of the segment through each point describes the accuracy of the ellipse orientation: $A_\theta:=1-\Delta\phi/(\pi/2)$, where $0\leq\Delta\phi\leq\pi/2$ is the minimal angle between the principal eigenvectors of the estimated and sample covariance matrices.

For any given test signal, two overlapping factors seem to reduce network performance: (i) a smaller chirp mass, since the information content (variability) of the signal is larger in this regime, and (ii) vicinity to the $(M_c,\eta)$ boundary. This latter effect is somewhat expected, since the posterior is closer to a truncated Gaussian in the case of a near-boundary signal, and the Gaussian defined by the sample covariance matrix actually tends to be a poorer fit than the (truncated) network estimate.

The distribution of network prediction errors for a test set of 5000 signals is shown in Fig.\ \ref{fig:meanstd}; the error in the mean value of each parameter $\Delta\mu(\theta)$ is quoted relative to the respective prior interval length $\Delta\theta$, while the error in each standard deviation is taken to be the quantity $\Delta\lg{\sigma(\theta)}$ (such that a value of $+1$ corresponds to an overestimate of $\sigma$ by a factor of ten). The network recovers the mean values at around $98\%$ accuracy, but can overestimate the standard deviations by up to a factor of three.

In Fig.\ \ref{fig:meanstd}, we also show the error distribution of $\mu(M_c)$ and $\sigma(M_c)$ for a second \textsc{Percival} network that estimates the one-dimensional posterior of $\theta=M_c$, for signals distributed according to the uniform prior
\begin{equation}
\label{eq:prior2}
p_2(\vartheta)\propto\mathbf{1}_{\Delta M_c\times\Delta\eta\times\Delta\chi^2\times\Delta\rho}(M_c,\eta,\chi_1,\chi_2,\rho),
\end{equation}
where $\Delta\chi=[-1,1]$, and the other intervals are given as before. This network is identical to the first, except that it outputs 18 quantities specifying a three-component bivariate Gaussian mixture (five for each component, plus three weights). Even though the family of posteriors over the full five-dimensional prior space is larger and more complex, the second network yields comparable results to the first, but does have some difficulty in resolving the multimodality that arises in certain posteriors.
\begin{figure}[!tbp]
\centering
\includegraphics[width=\columnwidth]{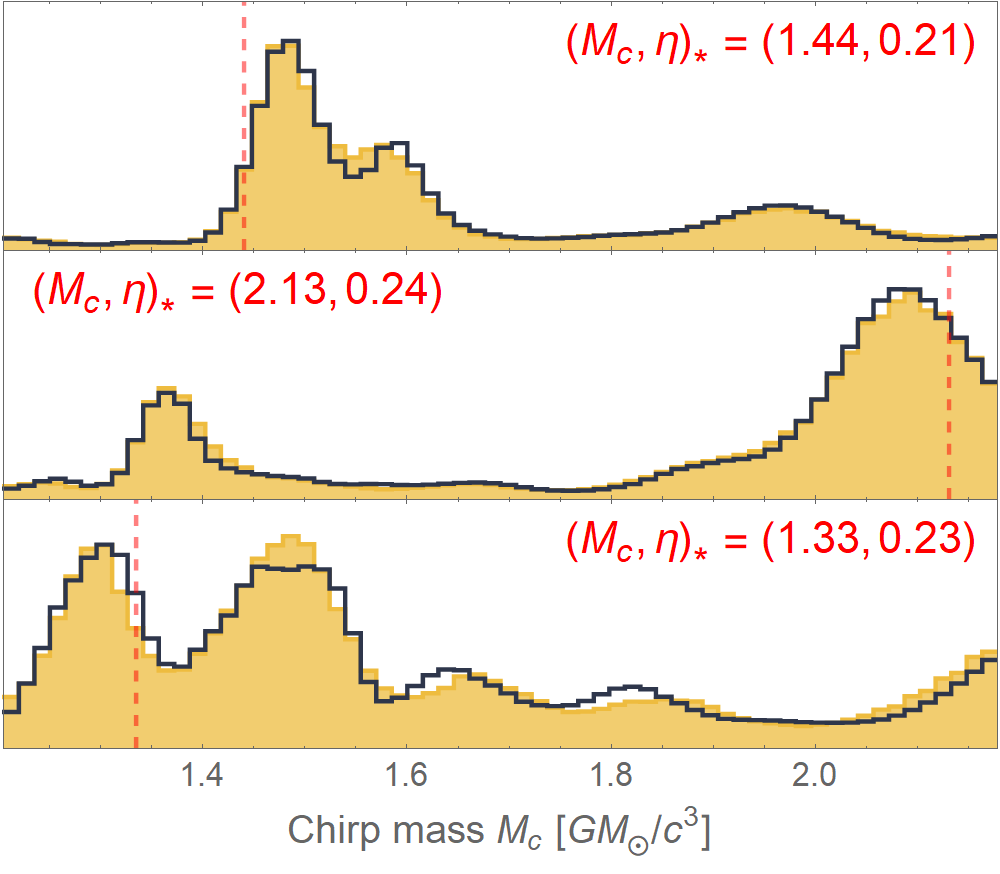}
\caption{Estimated (yellow) and true (black) $M_c$ posteriors for three weak test signals (red), with the prior $p_3(\vartheta)$.}
\label{fig:lowsnr}
\end{figure}

To show the viability of the histogram representation, we present results for a similar network (but with a \emph{softmax} output layer \cite{10.1007/978-3-642-76153-9_28}) that is trained on weak signals ($\rho=2$), which exhibit posteriors with complex features. The parameter of interest is $\theta=M_c$, and the prior is
\begin{equation}
\label{eq:prior3}
p_3(\vartheta)\propto\mathbf{1}_{\Delta M_c\times\Delta\eta}(M_c,\eta)\,\delta(\chi_1)\,\delta(\chi_2)\,\delta(\rho-2),
\end{equation}
where $\Delta M_c,\Delta\eta$ are given as before. While such a prior is not very relevant for gravitational-wave astronomy (since inference is unlikely to be conducted on sub-threshold sources), the \textsc{Percival} network is nevertheless able to estimate the highly nontrivial posteriors, due to the greater degrees of freedom in the posterior representation and the smaller prior space. Posteriors for three representative test signals are displayed in Fig.\ \ref{fig:lowsnr}.

\paragraph{Discussion.}

In this Letter, we give a proof-of-concept demonstration (\textsc{Percival}) of instantaneous Bayesian posterior approximation in gravitational-wave parameter estimation, using straightforward perceptron networks that are trained on large signal + noise sets drawn from the assumed parameter priors and noise sampling distribution. The computational cost of parameter-space exploration is shifted from the analysis of each observed data stream to the offline network-training stage, making this scheme useful for low-latency parameter estimation, or for the science-payoff characterization of future experiments. Notably, the loss function does not require likelihood evaluations, but only the true parameter values of the training examples. Thus, our scheme can be used whenever the likelihood is expensive or unknown, but forward modeling is efficient and we have access to many samples of noise. This classifies it as a likelihood-free inference method (see, e.g., \cite{NIPS2016_6084}), and distinguishes it from techniques that aim to emulate the likelihood \cite{10.1111/j.1365-2966.2011.20288.x}.

In our examples, we leverage the hyper-efficient \textsc{Roman} forward modeling \cite{PhysRevLett.122.211101}, which allows us to train networks over effectively infinite training sets. We find that relatively modest network architectures are able to approximate the quasi-Gaussian posteriors obtained for stronger signals, as well as the multimodal posteriors that occur at very low signal-to-noise ratios. We fully expect the accuracy of approximation to improve with network capacity and training iterations. Larger and deeper networks should also learn posteriors across broader regions of parameter space, although it may prove expedient to train separate networks for different regions.

The real-world analysis of gravitational-wave signals involves a number of complications that are not represented in our demonstration, such as multichannel-response data sets, the presence of \emph{extrinsic} parameters that describe the relative spacetime location of source and detector, and variations (or estimation error) in the noise spectral density. These are beyond the scope of this Letter, but can be handled by a combination of strategies: using more realistic and suitably parametrized waveform models to reduce degeneracy, designing networks with symmetries that make them insensitive to new degrees of freedom, and transforming (e.g., time-shifting, or whitening) the input data. Last, since convolutional neural networks have been successfully trained to recognize gravitational waveforms represented as strain time series, it should also be possible to combine these with a posterior-generating stage analogous to \textsc{Percival}, for a direct mapping from detector data to Bayesian posteriors.

\paragraph{Acknowledgements.}
We are grateful to Chad Galley for sharing his expertise in reduced-order modeling, and to Chris Messenger for correspondence on his related work. We also thank Milos Milosavljevic, Ian Harry and Christopher Berry for useful comments on the manuscript, and Natalia Korsakova and Michael Katz for helpful conversations. Finally, we acknowledge feedback from fellow participants in the 2018 LISA workshop at the Keck Institute for Space Studies. This work was supported by the Jet Propulsion Laboratory (JPL) Research and Technology Development program, and was carried out at JPL, California Institute of Technology, under a contract with the National Aeronautics and Space Administration. \copyright\,2020 California Institute of Technology. U.S. Government sponsorship acknowledged.

\bibliography{main}

\end{document}